# scientific reports

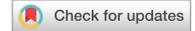

### OPEN
# Magnetic field amplification driven by the gyro motion of charged particles

Yan-Jun Gu✉ & Masakatsu Murakami

Spontaneous magnetic field generation plays important role in laser-plasma interactions. Strong quasi-static magnetic fields affect the thermal conductivity and the plasma dynamics, particularly in the case of ultra intense laser where the magnetic part of Lorentz force becomes as significant as the electric part. Kinetic simulations of giga-gauss magnetic field amplification via a laser irradiated microtube structure reveal the dynamics of charged particle implosions and the mechanism of magnetic field growth. A giga-gauss magnetic field is generated and amplified with the opposite polarity to the seed magnetic field. The spot size of the field is comparable to the laser wavelength, and the lifetime is hundreds of femtoseconds. An analytical model is presented to explain the underlying physics. This study should aid in designing future experiments.

Magnetic fields are detected throughout the universe and widely participate the astrophysical dynamics. Various fundamental phenomena, including coronal mass ejections[1,2], solar flares[3–5], closure of the planetary magnetosphere[6–8], $\gamma$-ray bursts[9–12] and pulsar winds[13–15], are dominated by variations in magnetic fields. Laboratory generation of strong magnetic fields has been an attractive research topic due to the significant role of magnetic fields in fusion instabilities and magnetized plasmas[16–21]. Coupling strong magnetic fields and high-power lasers supports cutting-edge research in laboratory astrophysics[22–24] and laser-driven charged particle accelerations[25,26].

Spontaneous magnetic field generation via laser-plasma interactions is well studied in decades. The Biermann battery effect[27], which generates a toroidal magnetic field based on electron motions along the density and temperature gradients in plasmas, was applied in early laser-driven laboratory astrophysical experiments[28,29]. With moderate laser intensities ($\sim 10^{15} \text{W/cm}^{-2}$) and a relatively long pulse length ($\sim$ ns), mega-gauss (MG) magnetic fields in plasmas at the edges of focal spots have been achieved. In the case of a high-intensity laser irradiating on a solid target, strong radiation pressure forms sharp density gradients, inducing the ponderomotive electric current. Numerical and theoretical studies predict a magnetic field strength of $\sim 100\,\text{MG}$[30–32] and the state-of-the-art experimental results are about $\sim 10\,\text{MG}$[33,34].

Although the mechanisms involved in the origination of magnetic fields in space are still uncertain, one of the widely accepted plausible scenarios is the turbulent dynamo, which amplifies the weak magnetic field[35–37]. Recent numerical studies and experiments with long pulse laser-produced colliding plasma flows have demonstrated the capability of seed fields amplification[38,39]. The concept of seed magnetic field amplification has also been applied in laser-plasma laboratories. Additionally, a microtube implosion for magnetic field generation has been proposed recently[40,41]. A strong magnetic field can be induced in the implosion process of the microtube structure via laser-plasma interaction based on a relatively weak seed magnetic field ($\sim 10$ MG). In particular, under some specific conditions, the generated magnetic field has the opposite direction to the seed magnetic field. While it is quite an interesting phenomenon, the microscopic physics remains unclear. Whether the electrons or the ions dominate field amplification has yet to be investigated. In this paper, we present the dynamic process of seed magnetic field amplification driven by the interaction of ultra-intense short laser pulses with microtube plasmas. Based on kinetic simulations by particle-in-cell (PIC) methods, the detailed kinetics are demonstrated. The results reveal that the direction of the produced magnetic field is anti-parallel to the seed magnetic field. The maximum amplitude of the induced monopole magnetic field saturates at several giga-gauss (GG) with a lifetime around hundreds of femtoseconds. Such studies provide a better understanding of the mechanism and insight for practical experiments.

Institute of Laser Engineering, Osaka University, Suita, Osaka 565-0871, Japan. ✉email: gu-yanjun@ile.osaka-u.ac.jp

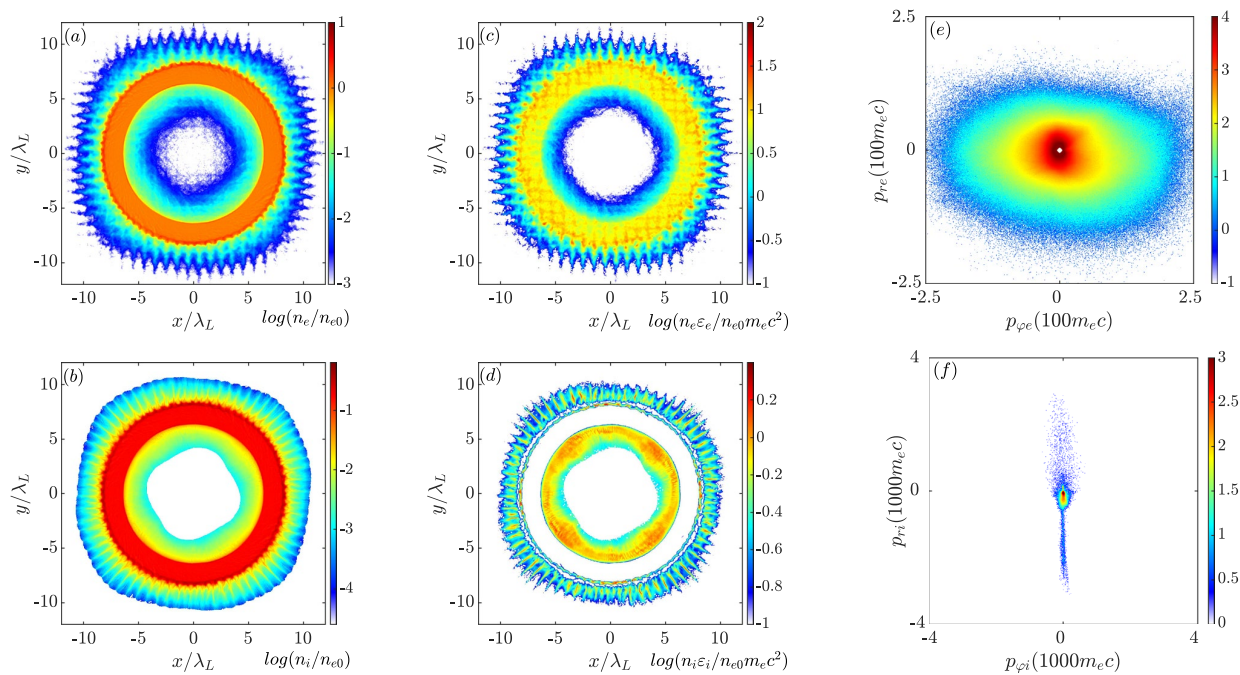

**Figure 1.** (**a**,**b**) are the density distributions of electrons and ions at t = 170 fs normalized to the initial electron density on a logarithmic scale. (**c**,**d**) are the corresponding distributions of energy density on a logarithmic scale. (**e**,**f**) are the electron and ion distributions in momentum space.

## Results

**Simulation setup.** Four linear-polarized Gaussian pulses with a peak intensity of $10^{21}$ W/cm$^2$ propagating along the $\pm x-$ and $\pm y-$axes are focused on the edge of the target. The normalized amplitude is $a_0 = eE_L/m_e\omega_L c \approx 48$, where $E_L$ and $\omega_L$ are the laser electric field strength and frequency, $e$ and $m_e$ are the electron charge and mass, respectively; $c$ is the speed of light in vacuum. The pulse duration is $\tau_L = 100$ fs. The laser wavelength is $\lambda_L = 0.8\,\mu$m. The initial carbon ion density is $n_{i0} = 3 \times 10^{22}$cm$^{-3}$ with a pre-ionized charge $Z = +6$ and an atomic mass number $A = 12$, which indicates the initial electron density is $n_{e0} = 6n_{i0} = 1.8 \times 10^{23}$cm$^{-3}$. In the two-dimensional (2D) case, the microtube target is set as a ring with an inner radius of $R_{\text{inner}} = 6\lambda_L$ and an outer radius of $R_{\text{outer}} = 9\lambda_L$. The simulation box has a size of 28 $\lambda_L \times$ 28 $\lambda_L$ in the $x-y$ plane. The mesh size for the 2D simulation is $\delta x = \delta y = \lambda_L/100$. The timestep is 0.006 $T_0$, where $T_0$ is the laser period. All the quasiparticles (420 per cell) are initially at rest. The seed magnetic field is uniformly distributed in the simulation box with the component out of the plane $B_z$, which satisfies the condition of $\nabla \cdot \mathbf{B} = 0$. Systematically, the cases with seed magnetic strength ranging from 10–60 MG are investigated. The corresponding magnetic field from the laser field is about $B_L = 2.89$ GG.

**Numerical results in two-dimensional case.** First, the main 2D case with the seed magnetic field $B_{z,0} = 60$ MG is presented. Here, all four laser pulses are p-polarized with magnetic field components of $B_z$. The longitudinally incident laser pulses have electric components of $E_y$ and the corresponding transversely incident laser pulses have electric field components of $E_x$. Figure 1a,b are the electron and ion density distributions when the peak intensity is irradiated on the target (t = 170 fs). A typical signature of Rayleigh-Taylor instability appears on the ablation surface. The spikes grow according to the direction of the explosions for both the electrons and the ions. The main part of the target keeps its original circular shape as seen in the densest part. The inner void, which is originally in vacuum, is partially filled with the hot imploded electrons and ions. Figure 1c,d present the energy density distributions on a logarithmic scale. The energy absorption of the electrons differs from that of the ions. For the electrons, most of the energy is dumped inside the target, which implies that the electrons are mainly heated by the laser irradiation. The energy density distribution of the ions displays a double-layer structure, which corresponds to an ion explosion and an implosion motion. In the middle of the two layers, the energy density is negligible, but the corresponding number density is high (Fig. 1b). Only the ions on the inner and outer surfaces under the implosion and explosion are accelerated as they experience a ponderomotive pressure and Coulomb potential. According to the ponderomotive scaling law, the electron temperature is estimated as $T_e = 44\sqrt{I_{L22}(\lambda_L/\mu m)^2} \approx 11$ MeV, where $I_{L22}$ is the laser intensity normalized to $10^{22}$W/cm$^2$. The electron temperature obtained from the numerical results is about 12 MeV, which is well consistent with the theoretical estimate. This also confirms that electron heating is dominated by the laser field. Here we define a Jacobian matrix, as

$$J_F(r,\varphi) = \begin{bmatrix} \frac{\partial x}{\partial r} & \frac{1}{r}\frac{\partial x}{\partial \varphi} \\ \frac{\partial y}{\partial r} & \frac{1}{r}\frac{\partial y}{\partial \varphi} \end{bmatrix} = \begin{bmatrix} \cos\varphi & -\sin\varphi \\ \sin\varphi & \cos\varphi \end{bmatrix}, \quad (1)$$

in which $x = r\cos\varphi$ and $y = r\sin\varphi$ according to the polar-Cartesian transformation. By employing

$$\begin{pmatrix} p_\varphi \\ p_r \end{pmatrix} = J_F \begin{pmatrix} p_y \\ p_x \end{pmatrix}, \quad (2)$$

the particle motion can be distinguished as the radial momentum, ($p_r$), and the azimuthal momentum, ($p_\varphi$). Figures 1e,f show the momentum space ($p_\varphi$, $p_r$) of the electrons and the ions. It should be noted that a positive radial momentum indicates a particle explosion and a negative radial momentum corresponds to an implosion. Similarly, the positive and negative azimuthal momenta correspond to the anti-clockwise and clockwise motion, where the clockwise and anti-clockwise refer to the center in the frame of reference ($r = 0$). The ions have a large radial momentum in both directions but relatively small $p_\varphi$, which means the ions are either exploding or imploding. Due to their heavier mass, the ions do not response to the fast oscillating components of the laser field. This is consistent with the energy density distribution. However, the electron momentum space has a broadened distribution along $p_\varphi$ and a relatively narrow distribution in the $p_r$ direction. It indicates that electron acceleration is dominated by the laser field.

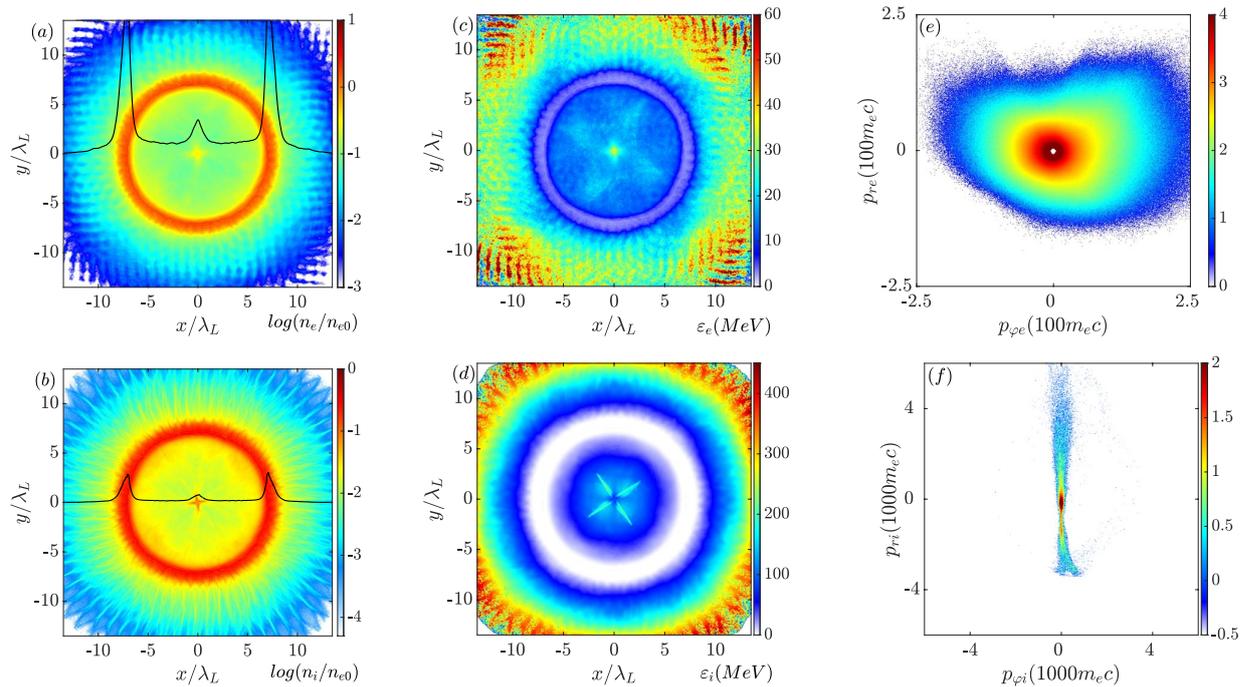

**Figure 2.** (**a**,**b**) are the density distributions of electrons and ions at t = 250 fs normalized to the initial electron density on a logarithmic scale. The black curves represent the linear density profiles along $y = 0$ normalized to $0.1n_{e0}$. (**c**,**d**) are the kinetic energy distributions of electrons and ions averaged in each mesh. (**e**,**f**) are the electron and ion distributions in momentum space.

When the imploded particles collapse into the center, they form a core and the local density increases significantly as seen in Fig. 2a,b, which is at t = 250 fs and the pumping energy from the laser pulses is almost terminated. The black curves represent the linear density profiles along $y = 0$ normalized to $0.1n_{e0}$. The corresponding charge density of the core can be estimated as $e(Zn_i − n_e) \approx 0.12en_{e0}$. Therefore, the central ions, which experience the Coulomb expulsive force, start exploding. In addition, part of the electrons attracted by the radial electric field are still imploding. Figure 2c,d present the kinetic energy distributions of the electrons and the ions averaged in each mesh. The imploded electrons around the core have a typical energy of about 35 MeV. Since the laser pulses cannot penetrate the target, the most energetic electrons, which are directly heated by the laser fields, are located on the outer surface. Comparing the momentum space shown in Figs. 1e and 2e, the azimuthal momentum of electrons becomes asymmetric, implying that a net spin current is formed. This net spin current is crucial to magnetic field amplification. In principle, the exploded and imploded ions have similar energies (Fig. 1f), in which the maximum positive and negative $p_r$ are close to each other. Once the ions collapse at the core, the Coulomb force decelerates and pushes them outwards. Therefore, the inner ions shown in Fig. 2d have a typical kinetic energy of around 200 MeV which is lower than that on the outer layers. This is also reflected in the momentum space in Fig. 2f. The exploded momentum greatly exceeds that of the imploded ones.

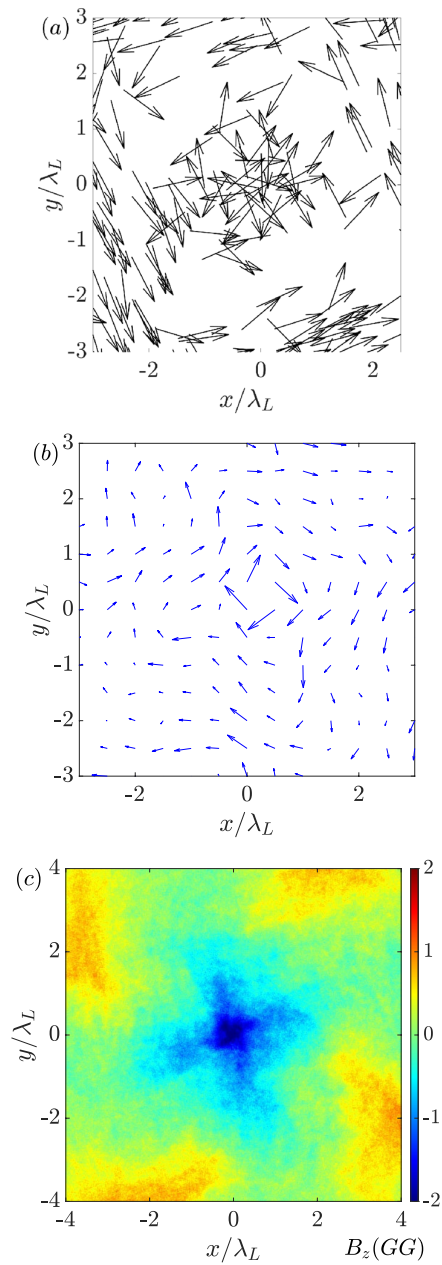

**Figure 3.** (**a**) Velocity vectors of the energetic electrons ($\varepsilon_e > 80$ MeV) at t = 250 fs around the core region. (**b**) is the corresponding current vectors. (**c**) The distribution of $B_z$.

The azimuthal momentum of electrons displays a preference of positive $p_\varphi$ (Fig. 2e). From the viewpoint of magnetic field amplification, the most significant contributions come from the hot electrons around the core region. The corresponding velocity vectors of the inner electrons, which have an anti-clockwise motion (Fig. 3a), are consistent with the momentum distribution, whereas the ion flows remain mainly along the radial direction with minor components in the azimuthal direction. Since the seed magnetic field is about 1% of the laser magnetic strength, it has a negligible effect on the ion motion. Therefore, the azimuthal current is dominated by the electron dynamics. A net spin current with a clockwise direction is formed in Fig. 3b. According to Ampere's circuital law, a magnetic field pointing in the negative z-direction is generated. As seen in Fig. 3c, the amplitude of the magnetic field reaches a few giga-Gauss with a radius around laser wavelength. Outside the core region, the magnetic field changes its sign. Although the total $B_z$ flux in the simulation box remains constant, a strong monopole region is induced.

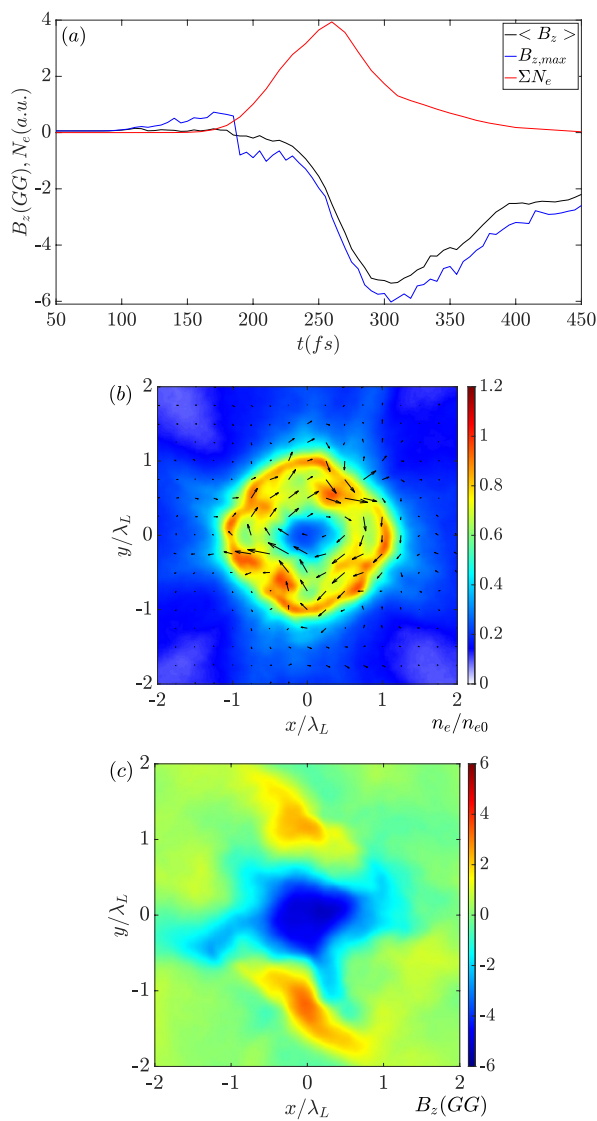

**Figure 4.** (a) Time evolution of the average (black) and maximum (blue) of $B_z$ in the core region ($\sqrt{x^2 + y^2} < \lambda_L$). Red profile indicates the total number of anti-clockwise rotating electrons in the core with an arbitrary unit. (**b,c**) are the distributions of linear electron density and $B_z$ at t = 300 fs. Black arrows in (**b**) are the corresponding current vectors.

As more electrons are injected in the later stage, the spin current and the corresponding magnetic field become stronger. Figure 4a plots the amount evolution of the electrons with a positive azimuthal momentum in the core region along with the evolution of average and maximum $B_z$ in the same region. The growth of the magnetic field amplitude is following the increase of electron number with a time delay of tens of femtoseconds. In the first 200 fs, magnetic field growth is insignificant. Then the magnetic field begins to increase when the energetic particles enter the center. The average and maximum $B_z$ amplitudes are close to each other, suggesting that the monopole magnetic region is quasi-uniform. The magnetic strength becomes saturated around 6 GG at t = 300 fs, which is more than double that of the laser magnetic field. It decays as the rotating electron number decreases. The induced monopole magnetic field has a lifetime of about 200 fs estimated by the full width half maximum (FWHM) of the evolution. Figure 4b shows the electron density distribution when $B_z$ reaches the peak at t = 300 fs. Consistent with the current vectors (black arrows), a clear high-density loop is formed. The vector length represents the current strength, and the current enhances the magnetic field. According to the distribution shown in Fig. 4c, the spot size of the monopole $B_z$ is slightly compressed, while the field amplitude remains relatively uniform.

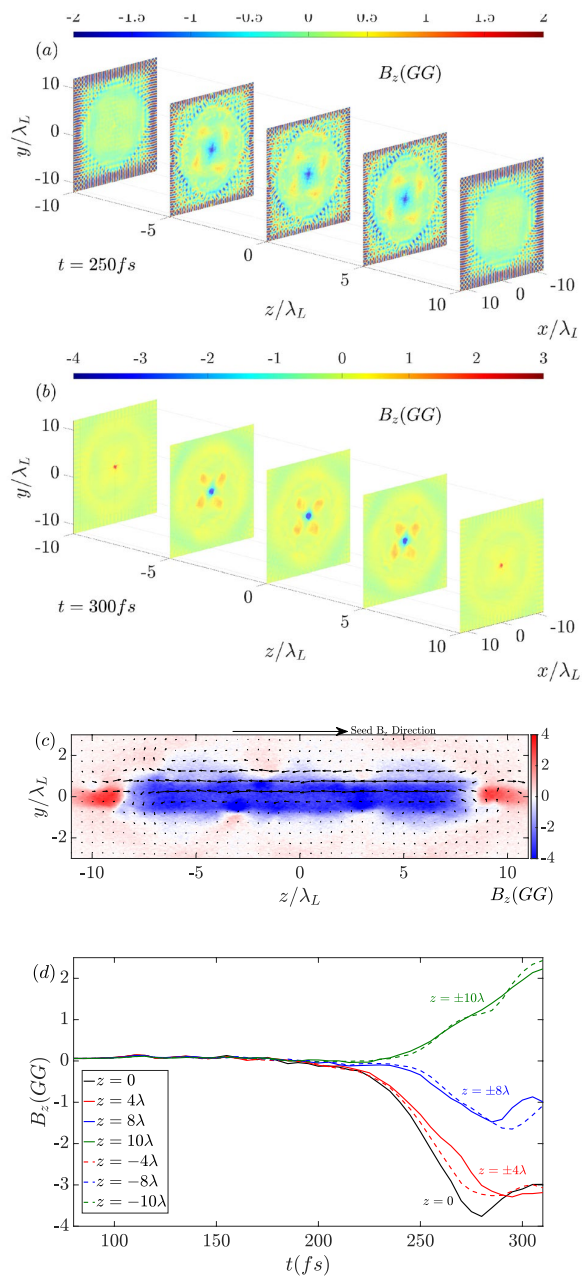

**Figure 5.** $B_z$ distributions at different slices along the $z$-direction ($z = 0, \pm 5\lambda_L, \text{ and } \pm 10\lambda_L$) at (**a**) t = 250 fs and (**b**) t = 300 fs. (**c**) Distribution of $B_z$ in the $y - z$ plane at t = 300 fs. (**d**) Time evolutions of $B_z$ at $z = 0, \pm 4\lambda_L, \pm 8\lambda_L, \text{ and } \pm 10\lambda_L$ along the central line.

**Numerical results in three-dimensional case.** To test the robustness of such a microtube structure, a three-dimensional (3D) simulation with the same parameters is conducted. In the $z$-direction, the length of the tube is set as 20 $\lambda_L$. Due to computational ability limitations, the mesh size for the 3D simulation is reduced to $\lambda_L/50$. Figure 5a,b show the distribution of $B_z$ in the different slices along the $z$-direction at t = 250 fs and t = 300 fs, respectively. Ideally, the 2D results can be regarded as the middle plane in a microtube with an infinite length. Comparing the central slices of $B_z$ ($z = 0$) with the 2D results presented above, the structures are well consistent except that the amplitudes of $B_z$ in 3D are slightly lower. The 3D effect is reflected on the outflow of the particles along the $z$-direction. According to the outflow, it can be predicted that the current strength for amplifying the magnetic field is weaker than that in the 2D case and the corresponding magnetic field also becomes weaker. The sign of the magnetic field changes at the boundaries of the tube ($z = \pm 10\lambda_L$). Figure 5c shows a slice of the $B_z$ distribution in the $y - z$ plane at t = 300 fs. The strong magnetic field covers a relatively large region with a length longer than 16 $\lambda_L$. From the experimental viewpoint, such a large area is feasible for diagnostics. The black arrows show the magnetic field vectors of $(B_y, B_z)$. The vectors are similar to the magnetic field lines of a solenoid. Figure 5d shows the time evolution of $B_z$ at different locations along the central line. The amplitude of $B_z$ peaks at the center $z = 0$ and gradually decreases on both sides. Similar to the 2D results, the amplitude of $B_z$ is also saturated around t = 300 fs, which implies that the 2D and 3D implosion dynamics are consistent with each other. The maximum $B_z$ in the 3D case is about 4 GG, which is about 25% lower than the 2D case but is much higher than the initial laser magnetic field.

### Particle dynamics and magnetic field polarity

An unclear issue is the direction of the amplified magnetic field. Here, it is in the opposite direction to the seed magnetic field. Intuitively, the imploded charged particles ($p_r < 0$) experiencing the positive seed magnetic field ($B_{z,0} > 0$) will be applied by an azimuthal Lorentz force ($f_\varphi = -qv_r B_{z,0}/c$). The ions obtain a positive azimuthal momentum ($f_{\varphi i} > 0$) while the electrons obtain a negative azimuthal momentum ($f_{\varphi e} < 0$). Therefore, a Larmor hole is formed in the center, as discussed in Ref.[40]. For an observer located inside a Larmor hole, it feels like the ions flow anti-clockwise and the electrons flow clockwise as depicted in Fig. 6a, in which the Lagrangian motions of charged particles are solved by the Boris algorithm. The initial condition is assumed to be a positive uniform seed magnetic field, and all the charged particles have a charge-to-mass ratio of $|q/m| = 1$. The net spin current helps form a positive magnetic field parallel to the seed $B_{z,0}$ inside the Larmor hole.

However, as the previous kinetic simulations presented, the situation is much complicate. Figure 2f implies that the ion azimuthal momentum is weak. Due to their heavy mass, the ions have a larger gyroradius, which is given as $r_{g,i} = m_i v_i c/ZeB$, and the corresponding deviation to the center is negligible. In other words, ion rotations do not contribute to the spin current and the amplification of the magnetic field. The imploded ions accumulate in the central core, which forms an unneutral charge region that results in a strong electrostatic field. The local electric field can be distinguished as a radial and an azimuthal field by applying the Jacobian matrix in Eq. (1)

$$\begin{pmatrix} E_\varphi \\ E_r \end{pmatrix} = J_F \begin{pmatrix} E_y \\ E_x \end{pmatrix}. \tag{3}$$

Figure 6b shows the corresponding $E_r$ obtained in the 2D simulation at t = 250 fs normalized by the laser electric field ($E_L$). The black curve is the profile of $E_r$ along $y = 0$, which is consistent with the density profiles shown in Fig. 2a,d according to $\nabla \cdot E_r = 4\pi e(Zn_i - n_e)$. The amplitude of the radial electric field is comparable to the laser electric field. It attracts electrons into the core region to balance the local electric charge. In this case, electron motions in the radial and azimuthal directions can be expressed as

$$\frac{dp_{re}}{dt} = -eE_r - \frac{e}{c}v_{\varphi e}B_z, \tag{4}$$

$$\frac{dp_{\varphi e}}{dt} = -eE_\varphi + \frac{e}{c}v_{re}B_z \approx \frac{e}{c}v_{re}B_z. \tag{5}$$

The azimuthal electric field is neglected since the ion azimuthal motion is weak. If the effect of $E_r$ is not taken into account, only the magnetic field affects the electrons. Then the kinetic energy of electrons is conserved as $p_{\varphi e}^2 + p_{re}^2 = const$, and the particles motions are shown in Fig. 6a. However, the appearance of a strong $E_r$ as shown in Fig. 6b changes the dynamics. Once the electrons enter the core region, they are accelerated by the radial Coulomb field and penetrate through the center. Then the imploding motion is transferred to the explosion. With the effect of the seed magnetic field ($B_{z,0} > 0$), the exploded electrons are bent by the Lorentz force ($\dot{p}_{\varphi e} = f_{\varphi e} > 0$), i.e., the electrons obtain the angular momentum $L_e = r_e \times p_{\varphi e}$. The radial Coulomb field provides a centripetal force to confine the rotating electrons with a trapping condition of

$$\gamma m_e v_{\varphi e}^2 < eE_r r_e, \tag{6}$$

where $\gamma$ is the Lorentz factor of the electrons. According to the numerical results as $E_r \sim E_L$ and $r_e \sim \lambda_L$, the binding potential energy is about 60 MeV. Electrons with the proper energy are trapped by the Coulomb potential and unable to leave the core region. The electron kinetics calculated by the Boris algorithm with a radial Coulomb field and seed magnetic field is shown in Fig. 7. A test electron is injected from the $-x$ direction and bent by the seed magnetic field. Once it enters the strong Coulomb field region ($x < 1$), it is trapped by the radial field and rotates to the $+x$ region instead of escaping towards the $+y$-direction. It then spins around the core in an anti-clockwise direction. According to the electron motion, a clockwise current forms and the corresponding negative magnetic field is generated.

To demonstrate this in the 2D PIC simulation, a bunch of electrons injected into the core region are traced. Figure 6c presents the corresponding trajectories. Once the electrons enter the core around 250 fs, they are confined in the region of $r \sim \lambda_L$, which is consistent with the spot size of the amplified magnetic field. The corresponding electrons spin around the core in an anti-clockwise direction. The momentum space of the electrons in the core region is shown in Fig. 6d. As predicted, most of the electrons have $p_{\varphi e} > 0$. Such a motion induces a clockwise spin current and produces a magnetic field in the opposite direction to the seed magnetic field $B_{z,0}$. The effect of the radial electric field also explains the magnetic polarity reversal at the boundaries of the microtube obtained from 3D simulations as shown in Fig. 5c. Because no ions implode into the region out of the tube, an accompanying strong radial electric field is not formed. In this case, the electron motions are driven only by the seed magnetic field, which is corresponding to the situation shown in Fig. 6a. The gyro rotation of the electrons forms a Larmor hole in the center, which results in an effective anti-clockwise current loop. The corresponding magnetic field is parallel to the seed magnetic field, as discussed in Ref.[40].

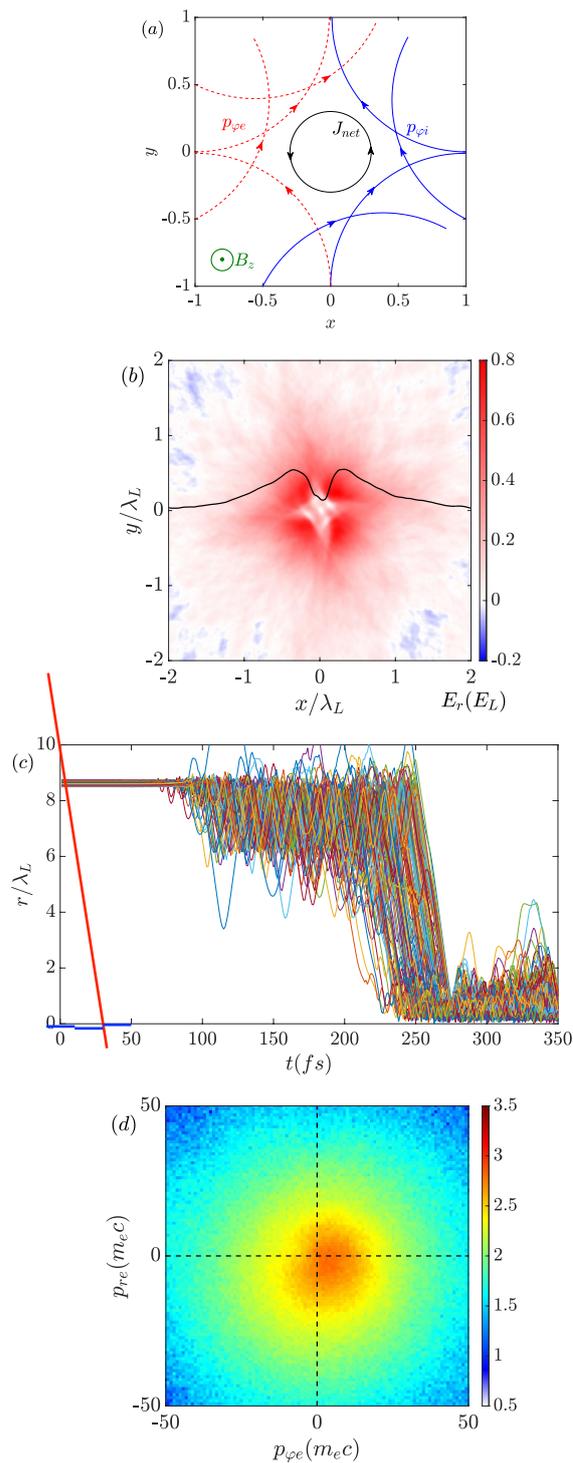

**Figure 6.** (**a**) Schematic of the imploded charged particles bent by the seed magnetic field before reaching the core. Trajectories are solved by the Boris algorithm with a positive uniform seed magnetic field. (**b**) Radial electric field around the core region obtained in the 2D simulation at t = 250 fs. (**c**) Trajectories of the trapped electrons. (**d**) Momentum space of the inner electrons at t = 250 fs.

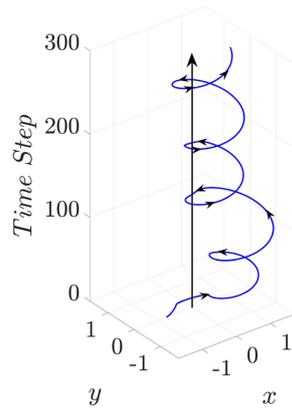

**Figure 7.** Electron trajectory solved by the Boris algorithm with a uniform seed magnetic field and a strong radial Coulomb field.

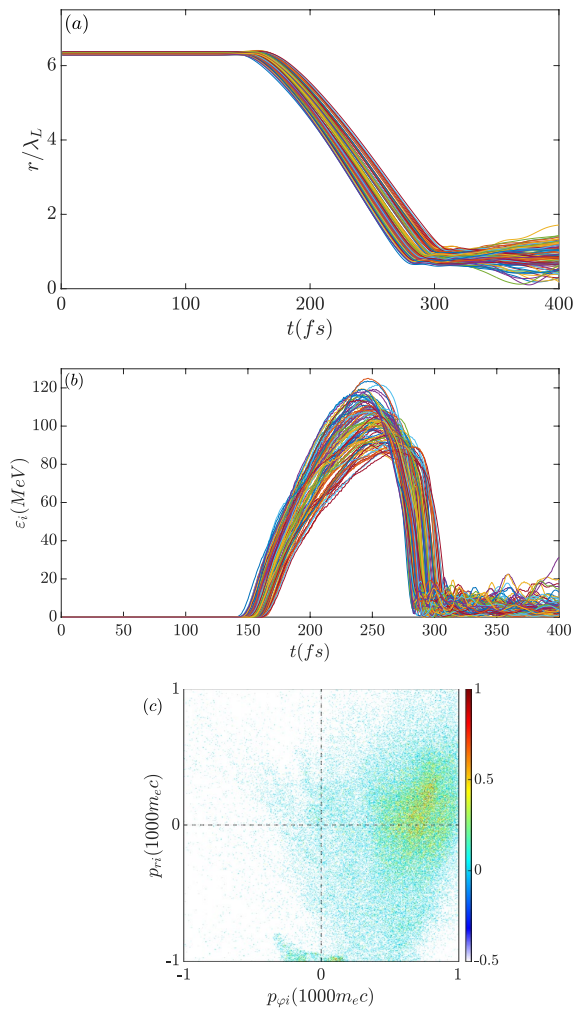

**Figure 8.** (**a**) Radius and (**b**) energy evolution of the characteristic imploded ions. (**c**) Momentum space of the inner ions at t = 350 fs.

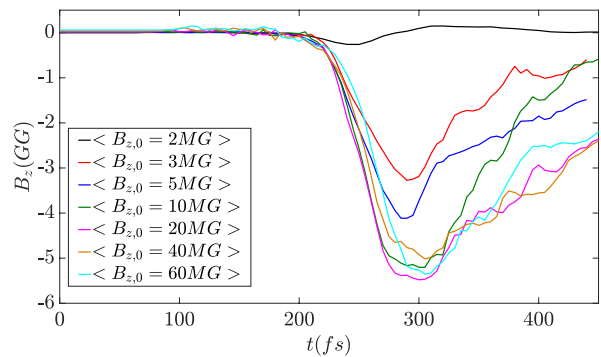

**Figure 9.** Evolutions of the average $B_z$ in the core region ($r < 1.5\lambda_L$) obtained in 2D simulations.

To discuss the effect of ion dynamics on the magnetic field variation, Fig. 8a,b show the characteristic ion trajectories and energy evolutions, in which the typical ions from the inner surface of the target are selected. The ions start imploding at t = 150 fs with the sound speed, $C_s = \sqrt{ZT_e/m_i} \approx 0.1c$ and gain energy. From 250 fs to 300 fs, the ions reach the core region and experience deceleration. Comparing with the magnetic field evolution shown in Fig. 4a, the maximum magnetic field strength is related to the end of ion implosion and energy loss. The radius of the ions grows again after 300 fs. Experiencing the growth of the induced magnetic field ($B_z < 0$) and the radial Coulomb field, the ions in the core region start exploding and rotating. According to the ion motion equations,

$$\frac{dp_{ri}}{dt} = ZeE_r + \frac{Ze}{c}v_{\varphi i}B_z, \qquad (7)$$

$$\frac{dp_{\varphi i}}{dt} \approx -\frac{Ze}{c}v_{ri}B_z, \qquad (8)$$

an explosion $v_{ri} > 0$ induces $\dot{p}_{\varphi i} > 0$. The phase space of the core ions at t = 350 fs (Fig. 8c) shows $p_{ri} > 0$ and $p_{\varphi i} > 0$ implying an explosion and anti-clockwise motion. Such a motion corresponds to an anti-clockwise spin current, which is opposite to the electron-driven spin current and reduces the induced $B_z$. This is also consistent with the decay of the magnetic field after 300 fs. Along with ion explosion, the radial Coulomb potential also decreases and the electrons are no longer confined. The induced magnetic field is sustained for another hundred femtoseconds till the core structure is completely inflated.

## Discussions and outlook

In summary, the magnetic field amplification process can be separated into the following three stages. The first is laser-driven implosion in which hot electrons and imploded ions gain energy from the pumping pulses. The second is the trapping of electrons in which the electron angular momentum is converted to induce a magnetic field. As the number of injected and trapped electrons increases, the induced magnetic field is amplified. The third stage is dissipation of the induced magnetic field in which the magnetic field energy is transferred to the angular momentum of the inner ions. The strength of magnetic field decays as the ions expand. Although the growth and amplification of the magnetic field are mainly dominated by the electron dynamics, the lifetime of the magnetic field is determined by the ion motion from collapse to explosion. Employing high Z and heavy ion materials may realize a longer lifetime. Comparing with the 3D simulation results shown in Fig. 5d, the magnetic field strength obtained in 2D simulations in Fig. 9 is higher. It is due to the confined particles in 3D case have a drift velocity along the z-direction, which will reduce the current amplitude in the transverse plane. Therefore, the magnetic fields in the 2D cases are somehow overestimated. However, as we discussed in the 3D results part, the difference is about 25%.

The results from the 2D simulations with different seed magnetic fields in Fig. 9 indicate that the maximum amplitudes and the time evolutions are similar when the seed magnetic field is large enough. The purpose of the seed magnetic field is to give an initial angular momentum to the imploded electrons for radial Coulomb field trapping. It is non-trivial to give an estimate of the minimum required seed magnetic field. Considering the hot electrons are guided by the seed magnetic field with the Lamor radius $R_L$, the collective motion of the hot electrons form a Lamor hole in the center with the radius of $R_H = \sqrt{R_0^2 - R_L^2} - R_L$, where $R_0$ is the initial radius of the target. When $R_0$ is comparable to $R_L$, the Lamor hole radius is approximately $R_H = R_0^2/(2R_L)$. Such the hole radius should be sufficiently larger than the Debye length in order to form a net spin current. Here we define a dimensionless parameter,

$$\zeta = \frac{R_H}{\lambda_D} = \frac{R_0^2}{2R_L \lambda_D}, \qquad (9)$$

in which $R_L = T_e/eB_{z,0}$ is the Lamor radius of the hot electrons and $\lambda_D = \sqrt{T_e/4\pi n_h e^2}$ is the Debyle length in the core region with the local hot electron density of $n_h$. For the purpose of magnetic field generation, the dimensionless parameter should satisfy

$$\zeta = \frac{e^2 \sqrt{\pi n_h} R_0^2 B_{z,0}}{T_e^{3/2}} >> 1, \qquad (10)$$

here the hot electron density $n_h$ and temperature $T_e$ are determined by the pumping laser amplitude. Based on the current parameters employed in the above simulations, it is obtained that the seed magnetic field should be $B_{z,0} > 2.5$ MG. The results shown in Fig. 9 presents a dramatic difference when the seed magnetic field transits from 2 MG to 3 MG. Therefore, in potential experiments, an extreme magnetic field strength is unnecessary as the state-of-art amplitude of 10 MG is sufficiently strong[33,34]. With the current laser facilities and the seed magnetic field preparation method, the giga-Gauss magnetic field generation with the proposed regime and setup is expectable.

## Methods

The simulations are performed with the relativistic electromagnetic code EPOCH[42,43] in 2D and 3D cases. The 2D simulation box has a size of 28 $\lambda_L \times$ 28 $\lambda_L$ in the $x - y$ plane. The corresponding mesh size is $\delta x = \delta y = \lambda_L/100$. The timestep is 0.006 $T_0$, where $T_0$ is the laser period. All the quasiparticles (420 per cell) are initially at rest. In the 3D simulation, the box size is X × Y × Z = 28 $\lambda_L \times$ 28 $\lambda_L \times$ 20 $\lambda_L$, sampled by 1400 × 1400 × 1000 cells. The free and thermal boundary conditions are applied in treating fields and particles, respectively.

## Data availability

The data that support the plots and findings of this paper are available from the corresponding author upon reasonable request.

### Acknowledgements
This work was supported by the Japan Society for the Promotion of Science (JSPS). Computational resources were provided by the Cyber Media Center of Osaka University. The EPOCH code was developed as part of the UK EPSRC funded projects EP/G054940/1.

### Author contributions
Y.J.G. carried out the simulations, analyzed the results, generated the figures, and wrote the bulk of the manuscript. M.M. discussed the physics and interpreted the results. Both authors contributed to the preparation of the manuscript.

### Competing interests
The authors declare no competing interests.

### Additional information
**Correspondence** and requests for materials should be addressed to Y.-J.G.

**Reprints and permissions information** is available at www.nature.com/reprints.

**Publisher's note** Springer Nature remains neutral with regard to jurisdictional claims in published maps and institutional affiliations.